\documentclass[twocolumn]{article}

\usepackage[T1]{fontenc}
\usepackage[utf8]{inputenc}
\usepackage{lmodern}
\usepackage{authblk}

\usepackage[english]{babel}
\usepackage{graphicx}
\usepackage[a4paper,margin=1in]{geometry}
\usepackage{xcolor}
\usepackage{wrapfig}
\usepackage{hyperref}
\hypersetup{hidelinks} 
\usepackage[round,authoryear]{natbib}
\usepackage{amsmath}
\usepackage{orcidlink} 

\title{Tunable Snapping and Rigid Foldability in the Mars Origami Pattern}

\author[1,2]{Menelaos Raptis\,\orcidlink{0009-0008-2226-5241}}
\author[1]{Thomas C. Hull\,\orcidlink{0000-0002-3215-9767}}

\affil[1]{Department of Mathematics and Statistics, Franklin \& Marshall College, Lancaster, PA 17604, USA}
\affil[2]{Department of Astrophysical Sciences, Princeton University, Princeton, NJ 08544, USA}

\affil[ ]{\texttt{mr0194@princeton.edu}; \texttt{thomas.hull@fandm.edu}}

\date{June 2026}

\begin{document}

\twocolumn[
\maketitle

\begin{abstract}
Origami-inspired metamaterials exploit the interplay between geometry and elasticity to achieve programmable mechanical responses. Yet the origin and tunability of snap-through instabilities in non-rigidly foldable patterns remain poorly understood. Here we show that the Mars tessellation, a degree-4 vertex origami pattern composed of alternating square and rhombic faces, is not rigidly foldable because the folding-speed ratios required for vertex compatibility cannot be propagated consistently across neighboring units. This geometric incompatibility forces the facets to bend during folding, giving rise to a reproducible snap-through discontinuity in the force--displacement curve with a mean force drop of about $92.6 \pm 5.5 \%$, marking a transition between metastable states. Laser scoring of additional diagonal creases, guided by strain-field simulations, enables continuous tuning of the snap magnitude. These results reveal a general mechanism by which geometric frustration can be harnessed to program multistability in thin-sheet metamaterials.
\end{abstract}

\vspace{2em}]



\section{Introduction}

Origami-inspired mechanical metamaterials transform flat crease patterns into a three-dimensional structure whose mechanical properties arise from the coupling of geometry and material response \citep{Schenk2013, EidiniPaulino2015, Brunck2016}.  
These systems often display unusual behaviors such as multistability, snapping, and tunable stiffness \citep{Waitukaitis2015, Wang2024}, which make them attractive for deployable devices, energy-absorbing components, and programmable materials.  
Understanding the origin of these behaviors is particularly important because they can be tuned not only by the crease pattern itself but also by fabrication choices such as material thickness, crease scoring, and unit-cell size.

Previous studies have shown that even patterns that appear to have a single geometric degree of freedom can undergo a \emph{snap-through}: an abrupt transition between metastable states when the energy landscape develops two wells separated by a barrier \citep{Silverberg2015,Bende2015,Fang2017,Yasuda2015,Lyu2021,Meloni2021}.  
For example, \citet{Silverberg2015} combined experiments with an energy-based model of the classic square twist crease pattern, measuring the bending response of the sheet and the stiffness of its creases, then minimizing the total energy under a geometric fold-compatibility constraint.  
Their results showed that as the plane angle $\phi$ crosses a critical value the energy landscape shifts from having one stable equilibrium to two, revealing the geometric origin of the crease pattern's snap-through as well as the role of hidden deformation models such as facet bending.

\begin{figure}[h!]
    \centering
    \includegraphics[width=0.5\textwidth]{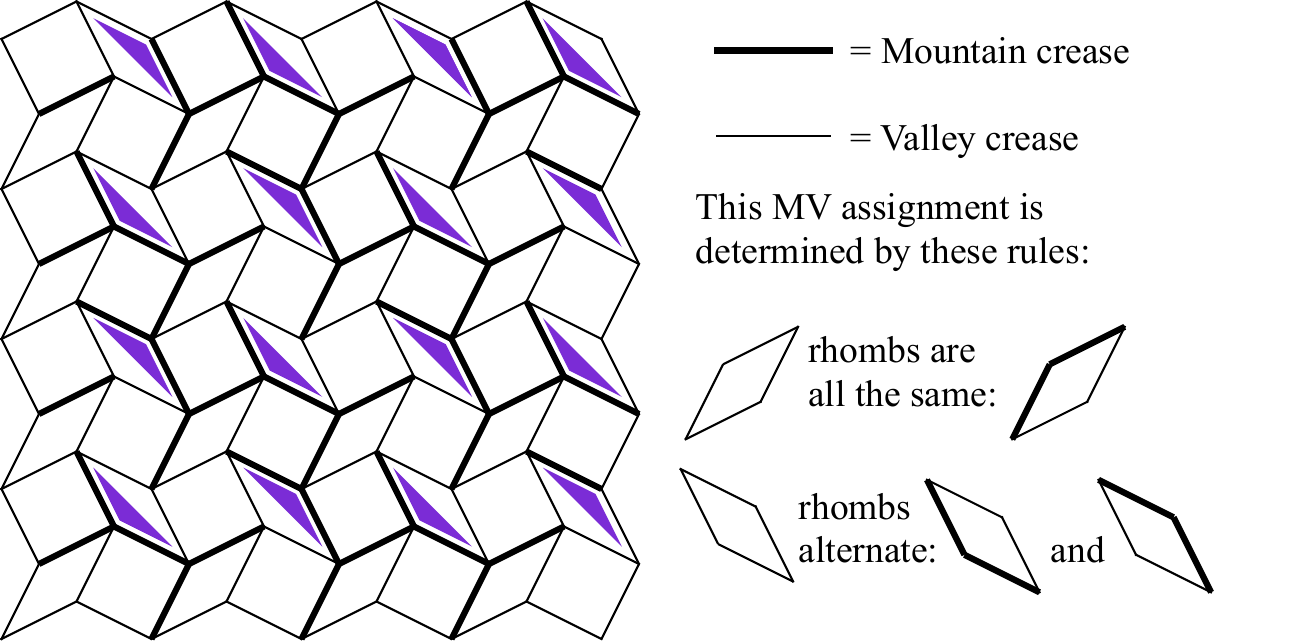}
    \caption{ 
    Barreto's Mars origami tessellation, with the specific mountain-valley (MV) assignment studied in this paper.}
    \label{fig:mars}
\end{figure}

The \emph{Mars pattern}, first introduced by \citet{Barreto1997} in the proceedings of the Second International Meeting of Origami Science and Scientific Origami, consists of alternating square and rhombic tiles joined by degree-4 vertices.
Barreto's paper presented the crease pattern and the mountain–valley assignment as a new folding motif of aesthetic and geometric interest, without offering a kinematic or mechanical analysis. The version of the pattern we study, shown in Figure~\ref{fig:mars}, can be viewed as a tessellation of square-twist units with a specific, fixed mountain--valley (MV) assignment that is propagated periodically across the sheet.

\begin{figure*}[h]
    \centering
    \includegraphics[width=.8\textwidth]{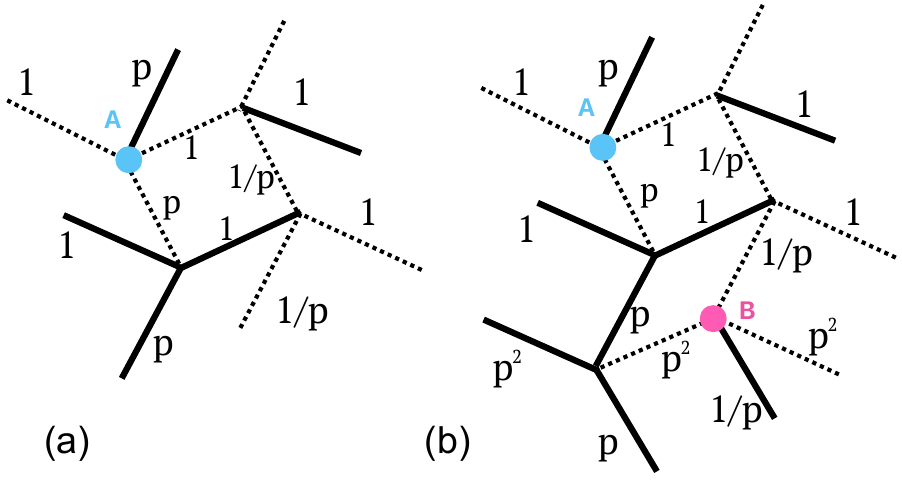}
    \caption{Folding-speed analysis of the Mars tessellation. 
    \textbf{Left:} The single square twist is consistent with folding speeds \(1\) (major) and \(p\) (minor). 
    \textbf{Right:} Propagating these speeds to neighboring vertices leads to an inconsistency at vertex B: the minor creases fold with speed \(1/p\) and the major with \(p^{2}\), violating the rigid-folding condition. 
    Mountain creases are indicated by solid lines, while valley creases are shown with dashed lines.}
    \label{fig:mars-rigid}
\end{figure*}

The mixed-vertex geometry from the MV assignment makes the Mars tessellation non-trivial from a mechanical point of view.  
At the level of a single degree-4 vertex, classical rigid-folding theory implies that the sector angles and MV assignment determine a unique set of folding-speed ratios along the incident creases \citep[Chapter 13.2]{hull2020origametry}.  
For an isolated square twist, which normally has all mountain creases around the square face, these constraints are incompatible around the central square, leading to isolated rigid configurations but no continuous rigid-folding path for the crease  pattern \citep{Silverberg2015}.  In our Mars MV assignment, however, the square twists each have one mountain and three valleys, or vice-versa, around the central square, and this does allow the square twist to rigidly fold continuously from the unfolded to the flat-folded state \citep{EvansLang2}. 
And yet, when these same square-twist-like units are assembled into the Mars tessellation, the vertex-compatibility conditions cannot be propagated consistently across neighboring units: the folding-speed ratios required at one vertex are incompatible with those required at adjacent vertices, as we show in Section \ref{sec:rigid_foldability}.  
As a result, the tessellation as a whole does not admit a global rigid-folding motion with perfectly flat facets. When the Mars pattern is realized in a thin elastic sheet, this geometric incompatibility is resolved by facet bending.  
Folding experiments show that out-of-plane deformations localize along specific diagonals of the square faces, allowing the structure to traverse configurations that are forbidden in the rigid-facet model.  
The measured force--displacement response exhibits a pronounced snap-through discontinuity, with a reproducible force drop separating two metastable configurations, indicating that geometric frustration at the level of crease kinematics is directly responsible for the observed mechanical bistability.

In this study, we combine vertex-compatibility analysis with controlled folding experiments to address two central questions:
(i) why the Mars tessellation fails to exhibit a continuous rigid-folding mode and how this geometric incompatibility manifests in the measured force–displacement response; and
(ii) how fabrication parameters --- such as crease scoring depth and the introduction of additional diagonal creases --- affect the magnitude and onset of the snap-through instability during folding.

This manuscript is organized as follows. In \S2, we analyze the rigid foldability of the Mars tessellation and show that the studied mountain--valley assignment is not rigidly foldable. In \S3, we describe the experimental methods used to fabricate and test the Mars patterns, including the force--displacement setup used to measure the response of the tessellation during folding and unfolding. In \S4, we present and discuss the experimental results. We first quantify the snap-through behavior of the baseline (\(3 \times 3\)) Mars pattern, which exhibits a reproducible force drop in the force--displacement curve. We then show that a larger (\(7 \times 7\)) Mars tessellation displays multiple snap-through events, suggesting that geometric incompatibility is relieved through localized transitions in extended patterns. Finally, we demonstrate that the snap-through response can be tuned by adding diagonal creases to the square faces and varying their scoring depth: increasing the scoring depth reduces the magnitude of the snap-through force drop. We conclude in \S5 with a summary of our findings and possible directions for future work.

\section{Rigid Foldability of the Mars Tessellation}
\label{sec:rigid_foldability}

Consider the vertex marked in blue in Figure~\ref{fig:mars-rigid}.  
We distinguish the two \emph{major} creases (opposite each other across this vertex with the same MV assignment) and the two \emph{minor} creases (the opposite pair with the opposite MV assignment).  
By classical results on rigid folding of degree-4 flat-foldable vertices \citep{EvansLang2,Lang2016Flasher}, the major creases fold faster than the minor ones. Moreover, the two major creases share a common folding angle, and likewise for the two minor creases. In fact, if $t_1$ is the major crease folding angle and $t_2$ the minor folding angle, then it can be shown \citep[Chapter 13.2]{hull2020origametry} that the ratio $p:=\tan(t_2/2)/\tan(t_1/2)<1$ is constant throughout the  folding motion of the vertex.

Normalize the folding speed of the parameterized major creases at the blue vertex to be~1; then the minor creases fold more slowly with constant parameterized speed \(0 < p < 1\), where \(p\) is determined by the degree-4 flat-foldability relation for the sector angles at the vertex. For the single square twist (Figure~\ref{fig:mars-rigid} (a)), these folding speeds are consistent around the square, so this isolated twist with this MV assignment is rigidly foldable. Extending the same MV assignment to the surrounding vertices in the Mars tessellation (Figure~\ref{fig:mars-rigid} (b)) produces an incompatibility. At the vertex marked in pink the minor creases fold with speed \(1/p\ (>1)\), while the major creases fold with speed \(p^{2}\).  
Thus, the ratio of major to minor folding speeds there is \(p^{3}\), instead of the ratio \(1/p\) required for a rigidly folding degree-4 vertex.

Hence the Mars pattern with this MV assignment is not rigidly foldable.  
This explains the observed snapping behavior: the incompatibility in crease-speed ratios forces the square faces to bend and store energy during folding.  
This observation also suggests that inserting diagonal creases into the square faces (changing the local vertex structure) could potentially restore rigid foldability to the tessellation, something we further explore below.

A more detailed analysis, presented in Appendix~\ref{app:rigid_analysis}, shows that when diagonal creases, which could be either mountain or valley, are introduced into the square  faces of our Mars tessellation in Figure~\ref{fig:mars-rigid} (b), the resulting structure still does not satisfy the conditions for infinitesimal rigid-foldability, as determined by the Jacobian and second-order conditions of the constraint system introduced by \citet{BelcastroHull2002}, and by the zero-area reciprocal diagram criterion and infinitesimal rigid-foldability conditions of \citet{WatanabeKawaguchi2009}, who showed that the existence of such a diagram is a necessary condition for rigid foldability. Therefore, even when square diagonal creases are added to this Mars pattern, the MV assignment remains non-rigidly foldable. It still stands to reason, however, that adding such diagonal creases could affect the snapping behavior of this crease pattern. 


\section{Methods: Snapping Behavior Analysis}
\subsection{Experimental setup}

\begin{figure}[!ht]
  \centering
  \includegraphics[width=0.3\textwidth]{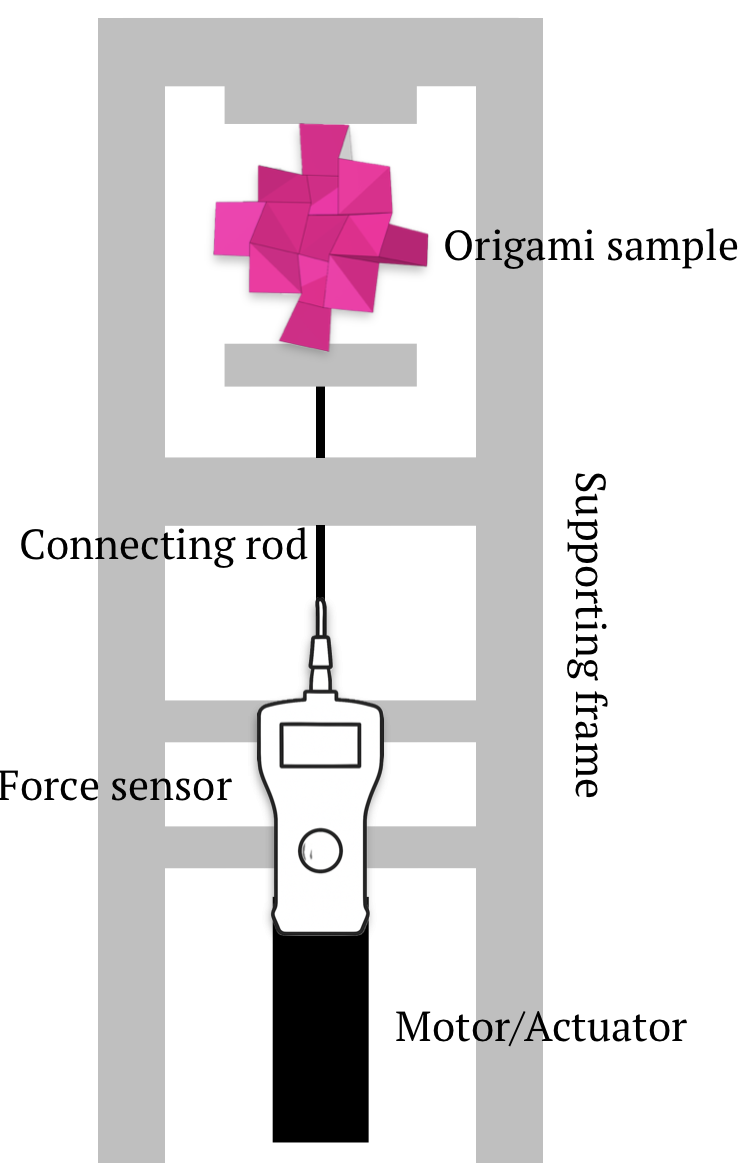}
  \caption{Schematic of the experimental setup showing the Mars origami specimen mounted vertically within the supporting frame and connected to the force sensor.}
  \label{fig:setup}
\end{figure}

The experimental setup used to measure the snapping behavior of the Mars origami pattern is shown in Figure~\ref{fig:setup}.
The origami sample is mounted vertically within a rigid supporting frame. 
One edge of the origami pattern is fixed to a stationary point on the supporting frame, while the opposite edge is attached to a connecting rod coupled to a Torbal digital force sensor (model FB5, 1~lbF~$\times$~0.0002~lbF). 
The force sensor is in turn connected to a motorized actuator that controls its vertical displacement. 
During each experiment, the actuator moves the force sensor downward in steps of 0.02~cm with a 0.1~s delay per step, corresponding to a total displacement of 6~cm.

\subsection{Experimental process}
\label{sec:experimental_process}

For the first experimental procedure (Section~\ref{exp1_results}), we generated the Mars pattern 31 times on 0.01~inch thick paper using a 
Glowforge laser cutter, a computer-controlled device that directs a high-powered laser beam to cut or engrave materials with high precision. After testing a range of settings, we selected a laser power of 48\% and a speed of 500 (out of a maximum of 3300) for cutting the boundary of the origami pattern, which produced clean edges without excessive burning. For crease formation, we used the same speed but reduced the power to 19\%, ensuring that the creases were etched into the paper without fully cutting through. Mountain (M) and valley (V) assignments were decided later by manual folding, which we acknowledge as a potential source of error in Section~\ref{sources_of_error}. The dimensions of the \(3 \times 3\) tessellation pattern were the same across all samples, with the central square of each square twist having a side length of 3 cm; this geometry was held fixed across all samples.

\begin{figure*}[h!]
    \centering
    \includegraphics[width=1\textwidth]{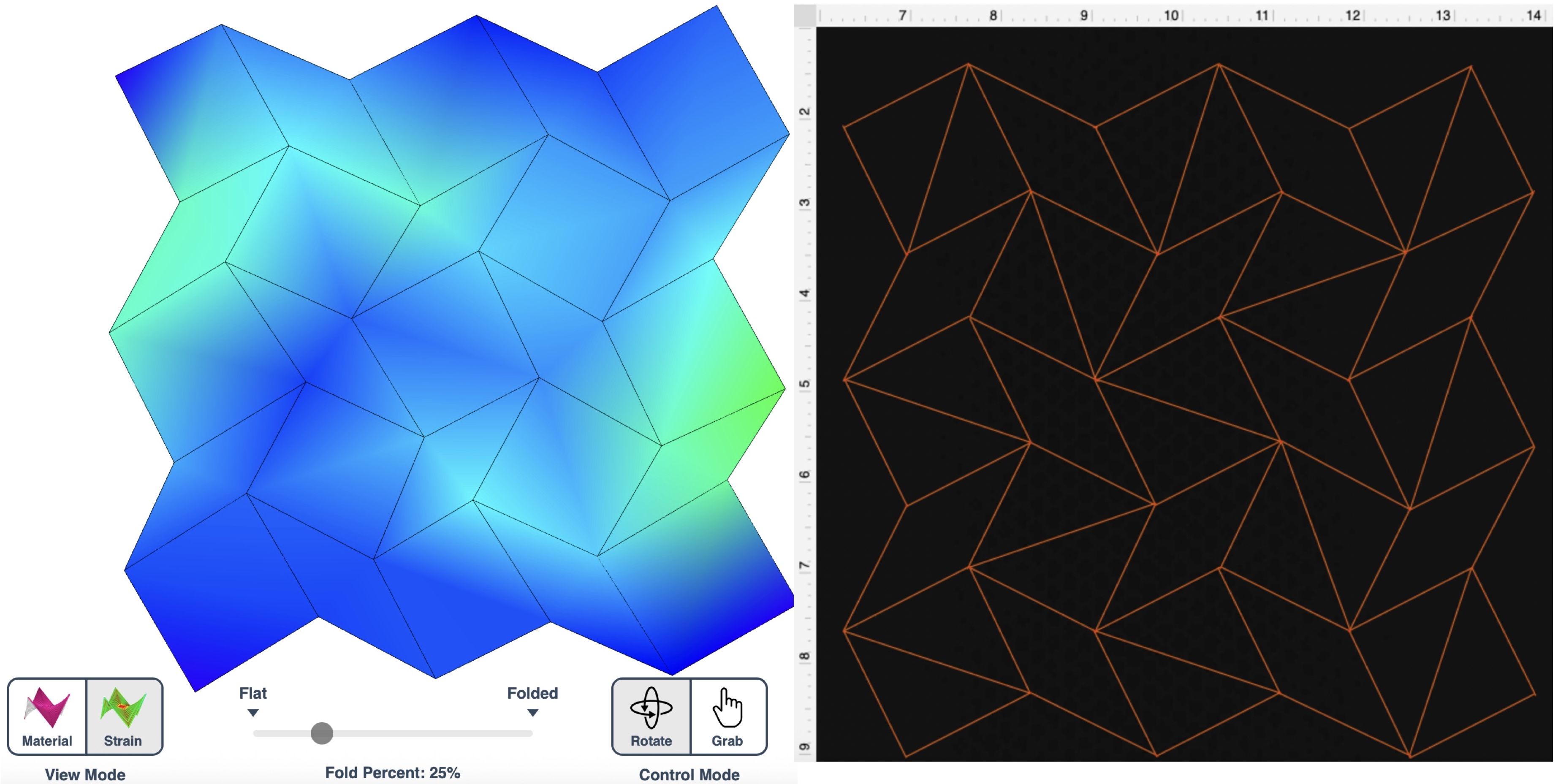}
    \caption{\textit{Left:} Axial strain visualization of the \(3 \times 3\) Mars tessellation generated using the Origami Simulator \citep{ghassaei2018fast}. Colors represent the magnitude of axial strain, ranging from blue (low strain) to green (high strain). Based on this visualization, additional diagonal creases were introduced along square faces to mitigate localized stress concentrations. \textit{Right:} The Mars pattern after incorporating these diagonal creases in the Glowforge software. While it may not be immediately apparent from the strain visualization which diagonals to add, each face was tracked throughout the folding process, revealing that, at least once, more intense strain occurred along one of the diagonals of each square.}
    \label{strain_visualization}
\end{figure*}

Then, in the second experimental procedure (Section~\ref{exp2_results}), in order to reduce and better understand the strain we measure from the machine, we used the \textit{Fast, Interactive Origami Simulation using GPU Computation} (\url{http://origamisimulator.org}) \citep{ghassaei2018fast}. This tool reformulates bar-and-hinge origami models into a compliant, explicit 
numerical solver that runs in parallel on the GPU. A crease pattern is first triangulated into a pin-jointed truss network, where each edge is modeled as a beam with linear-elastic axial constraints, creases are represented as torsional springs driving fold angles, and additional face constraints suppress in-plane shearing. The equations of motion are integrated explicitly with forward Euler time-stepping, with damping applied for stability, enabling rapid simulation without assembling a global stiffness matrix. For strain visualization, the software computes the engineering strain of each beam, 
\(\varepsilon = (\ell-\ell_{0})/\ell_{0}\), averages the absolute strain values at each vertex, and interpolates these values across faces. The resulting scalar field is mapped to a blue--red color spectrum (blue = low strain, red = high strain), providing immediate feedback on localized deformation during folding. Figure~\ref{strain_visualization} shows our \(3 \times 3\) Mars tessellation pattern, folded slightly and with strain as visualized in the Origami Simulator.

Guided by the strain-field visualization, we introduced additional diagonal creases along the square faces where the simulation indicated extrema in diagonal strain, with the aim of reducing the snapping behavior. These supplementary creases were engraved using the Glowforge laser cutter at precision power settings of 2, 4, 8, 10, and 12\%. For each setting, we repeated the folding–unfolding cycle and measured the corresponding percentage drop in applied force at the snap point. The relationship between the force-drop percentage and the laser power of the added creases is shown in the left panel of Figure~\ref{fig:adding_more_creases} (Section~\ref{exp2_results}) In this work, the Origami Simulator is used as a qualitative design and visualization tool rather than as a quantitatively predictive model; only relative spatial variations in strain are interpreted, not absolute strain magnitudes.

\begin{figure*}[h]
    \centering
    \includegraphics[width=1\textwidth]{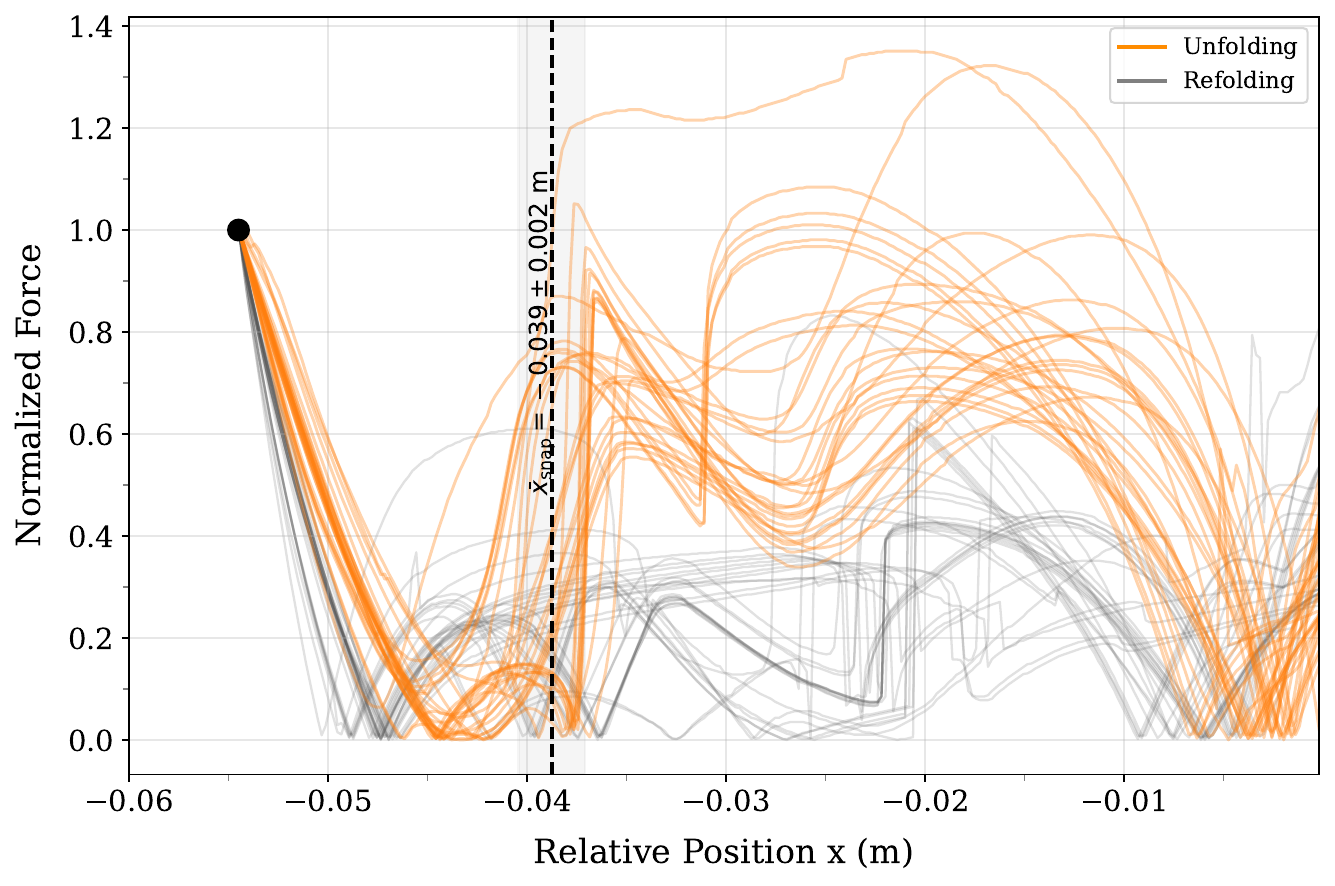}
    \caption{Normalized force measured by the load cell as the Mars origami pattern is unfolded (orange lines) and refolded (gray lines) as a function of relative position. Each curve corresponds to an independent experimental run on a distinct specimen fabricated with identical laser parameters, for a total of 31 Mars-pattern samples. The vertical dashed black line marks the mean snap-through position, \(x_{\mathrm{snap}}=-0.039~\mathrm{m}\), while the shaded band indicates the corresponding \(1\sigma\) variability (\(\pm 0.002~\mathrm{m}\)) across experiments, arising from the uncertainties discussed in Section~\ref{sources_of_error}.}
  \label{fig:main_mars_pattern}
\end{figure*}

\section{Results and Discussion}
\subsection{Force–Displacement Characteristics of the $3\times3$ Mars Pattern}
\label{exp1_results}

For the baseline \(3\times3\) Mars pattern, we performed \(N=31\) independent runs using thirty distinct paper specimens fabricated with identical laser parameters, as described in the first experimental procedure of Section~\ref{sec:experimental_process}.  

The orange curves in Figure~\ref{fig:main_mars_pattern} correspond to the unfolding phase of the pattern, while the gray curves represent the refolding (reverse) phase.  
The percentage drop in force, defined as \(\Delta F / F_{\mathrm{pre}} = (F_{\mathrm{pre}} - F_{\mathrm{post}})/F_{\mathrm{pre}} \times 100\%\), quantifies the magnitude of the snap-through discontinuity and is found to be: $92.6 \pm 5.5 \%$. Moreover, the average snapping position, defined as the midpoint between the local maximum and the subsequent trough corresponding to the largest snap-through event in each run and then averaged across all experiments, was found to be $\bar{x}_{\mathrm{snap}} = -0.039 \pm 0.002~\mathrm{m}$, with the snapping positions clustering within approximately $\sigma_x / |\bar{x}_{\mathrm{snap}}| \approx 5.1\%$ of the mean location. This relatively small fractional variation indicates a high degree of reproducibility across independently fabricated specimens when positioned under identical conditions in the experimental setup. We also note the occasional presence of a potential secondary, weaker force drop at a displacement of approximately $x \approx -0.02~\mathrm{m}$; however, this feature is not consistently observed across specimens and exhibits a substantially smaller and less abrupt force decrease than the primary transition. For this reason, the present analysis focuses on the dominant snap-through event.

Unlike the \(3\times3\) pattern, which exhibits a single snap-through event, the larger \(7\times7\) tessellation undergoes a sequence of localized snap-through transitions. As we analyze in Section~\ref{sec:7times7}, each force drop corresponds to the release of elastic energy stored in a subset of unit cells, rather than to a global rearrangement of the entire structure. This behavior indicates that geometric frustration in the Mars pattern acts locally, allowing different regions of the sheet to resolve incompatibility at different stages of the unfolding process.

\begin{figure*}[h!]
    \centering
    \includegraphics[width=1\textwidth]{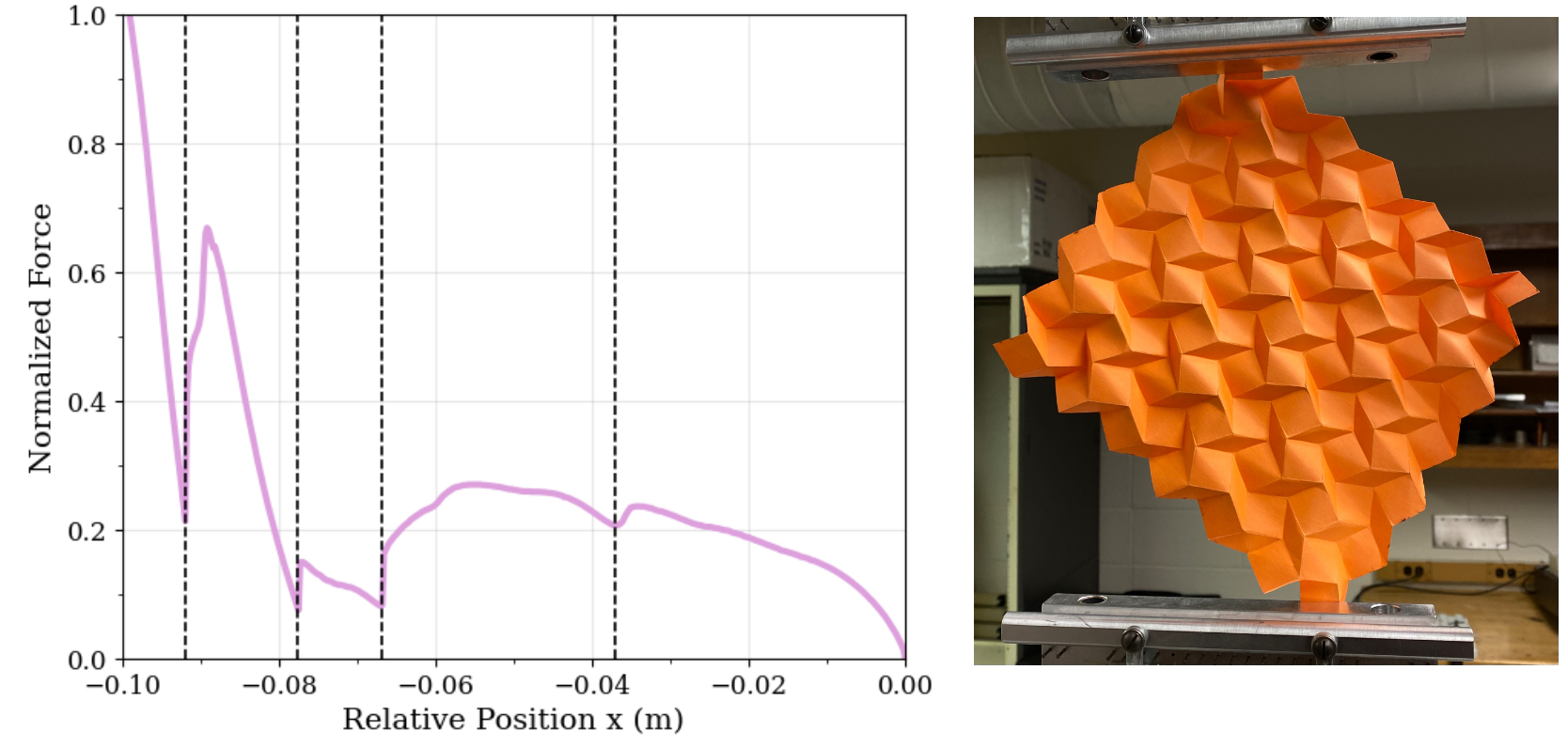}
    \caption{
    \textit{Left:} Normalized force versus relative position for the \(7\times7\) Mars tessellation pattern (purple curve).  
    Multiple snap-through events are visible as abrupt force drops of varying magnitude, ranging from approximately \(20\%\) to \(70\%\) in \(\Delta F / F_{\mathrm{pre}}\).  
    The corresponding snapping positions are indicated by vertical dashed black lines.  
    \textit{Right:} Photograph of the \(7\times7\) Mars origami sample during unfolding.}
    \label{fig:large_mars_pattern}
\end{figure*}

\subsubsection{The \(7\times7\) Mars Pattern}
\label{sec:7times7}
To further investigate how the snapping behavior scales with system size, we tested a larger \(7\times7\) Mars tessellation pattern fabricated with the same overall sheet dimensions as the \(3\times3\) baseline.  
Unlike the \(3\times3\) pattern, which exhibits a single snap-through event, this larger structure exhibited multiple distinct snap-through events during its unfolding cycle, each corresponding to localized collapses of sub-regions within the tessellation. The normalized force–displacement curve for the \(7\times7\) pattern (shown in Figure~\ref{fig:large_mars_pattern}, left) reveals several discrete force drops, ranging approximately from \(20\%\) to \(70\%\) in \(\Delta F / F_{\mathrm{pre}}\).  
The positions of these snapping events are marked by the vertical dashed black lines in Figure~\ref{fig:large_mars_pattern}. An optical image of the \(7\times7\) specimen at one of the critical configurations is shown on the right of the figure.

As observed from the indicated percentage drops in force in the left panel of Figure~\ref{fig:large_mars_pattern}, the percentage drop generally increases as the pattern unfolds, with the maximum drop occurring at the final snap, just before the pattern reaches its near-flat state. Given this observation, we now cautiously provide a physical explanation. Each force drop corresponds to the release of elastic energy stored in a subset of unit cells, rather than to a global rearrangement of the entire structure, with different regions of the sheet resolving incompatibility at different stages of the unfolding process. The physical origin of the increasing snap magnitude can be understood in terms of energy redistribution across the tessellation. During the early stages of unfolding, when many regions remain highly nonplanar, elastic energy released by a local snap can be partially absorbed by neighboring regions that have not yet reached their own instability, through additional facet bending and small rotations at the creases. As unfolding proceeds and the structure approaches the flat configuration, the number of regions capable of accommodating such redistribution progressively decreases: facets become nearly coplanar, bending degrees of freedom are limited, and the remaining incompatible regions are more strongly constrained. Consequently, the final snap-through event occurs when little capacity remains for local energy absorption, forcing a larger fraction of the stored elastic energy to be released globally, resulting in the largest observed force drop.

\begin{figure*}[h!]
    \centering
    \includegraphics[width=1\textwidth]{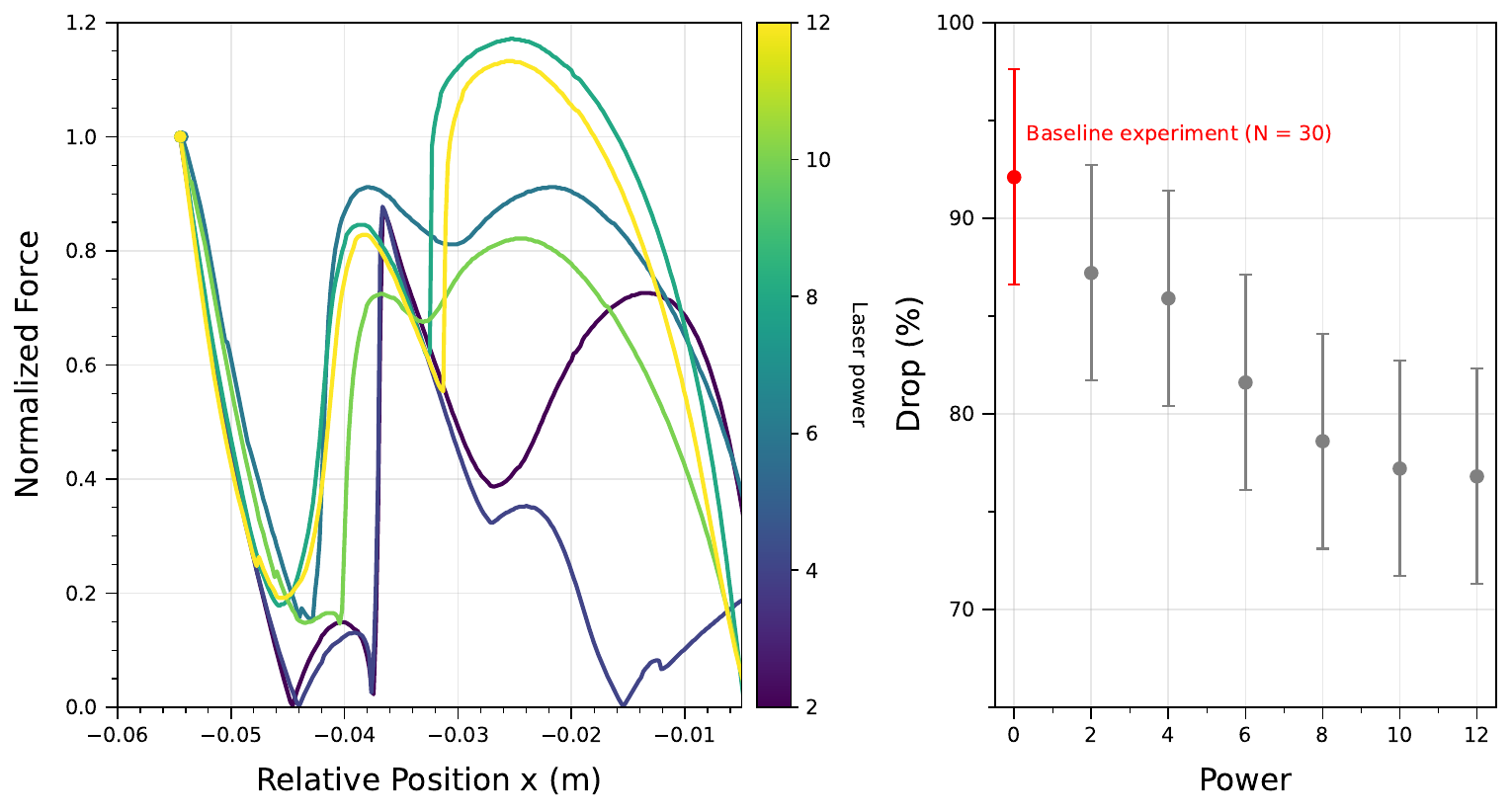}
    \caption{\textit{Left:} Normalized force (normalized by the force measured at the greatest displacement position of each run) as the Mars origami pattern is unfolded for different laser powers applied to the additional diagonal creases introduced on the square faces. Curves are color-coded by the applied laser power and plotted against the relative displacement. \textit{Right:} Magnitude of the dominant snap-through discontinuity, expressed as the percentage force drop, plotted as a function of the laser power used for the additional diagonal creases. The red point corresponds to the baseline experiment without additional laser-creased diagonals ($N=31$ independent specimens), while gray points correspond to individual laser-power conditions; error bars denote $\pm1$ standard deviation across specimens. These results demonstrate that the snapping behavior of the Mars pattern can be actively tuned: higher laser power, corresponding to deeper additional diagonal creases, leads to a reduced percentage drop in force at the snap-through transition.}

    \label{fig:adding_more_creases}
\end{figure*}

\subsection{Adding extra creases}
\label{exp2_results}
Following the methodology outlined in Section~\ref{sec:experimental_process}, we analyze the effect of introducing additional diagonal creases to the square faces of the Mars tessellation. These auxiliary creases are guided by the strain maps obtained from the GPU origami simulation and are engraved using the Glowforge laser cutter at power levels ranging from 2\% to 12\%, without assigning a specific M or V polarity.  
The goal of this series of tests is to examine how the stiffness of these added creases influences the magnitude of the snap-through discontinuity, expressed as the percentage drop in measured force during folding.

Figure~\ref{fig:adding_more_creases} (left) shows the force–displacement curves \(F(x)\) for the various crease powers, color-coded by laser setting. Each force–displacement curve corresponds to a single specimen per laser power.  
All curves exhibit a pronounced discontinuity associated with the snap event, consistently occurring at \(x_{\mathrm{snap}} \approx -0.04~\mathrm{m}\).  
This invariance in \(x_{\mathrm{snap}}\) across all specimens suggests that the onset of instability may not depend on the specific stiffness of the creases.  
The right panel of Figure~\ref{fig:adding_more_creases} summarizes how the magnitude of the snap-through discontinuity depends on crease stiffness.  
The percentage drop in force, \(\Delta F / F_{\mathrm{pre}}\), is plotted as a function of laser power \(P\).  
A clear monotonic decrease in \(\Delta F / F_{\mathrm{pre}}\) with increasing \(P\) is observed, indicating that deeper crease scoring lowers the energy barrier between metastable configurations and thereby suppresses abrupt snapping.

To determine whether the observed decrease in snap-through drop with increasing laser power is robust to measurement uncertainty, we performed a Monte-Carlo uncertainty propagation analysis. For each power level, the measured drop is modeled as a normal distribution with mean equal to the experimental value and standard deviation $\sigma=5.5\%$. We generated $10^{4}$ synthetic datasets by sampling from these distributions and computed the Spearman rank correlation coefficient $\rho$ between laser power and snap-through drop for each dataset. The probability of a decreasing association was estimated as the fraction of realizations for which $\rho<0$, yielding $P(\rho<0)=0.99$, indicating that approximately $99\%$ of uncertainty-consistent datasets exhibit a decreasing trend.

We repeated the same experimental process at several different specimen orientations to ensure that the observed trend was not direction-dependent. Although the individual force–displacement curves and exact drop percentages varied among orientations, the data consistently revealed a tight anti-correlation between the percentage drop in force and the laser power of the added creases.

\subsection{Sources of Uncertainty}
\label{sources_of_error}

Although the experiments were carefully designed and repeated under controlled conditions, several sources of uncertainty may influence the quantitative measurements and their interpretation.

First, small variations in the Glowforge laser cutter’s delivered power and focal depth can alter the effective stiffness of the scored creases. Even with identical settings, minor fluctuations in beam energy or surface reflectivity introduce local differences in thermal damage and residual stress, slightly changing how each crease behaves mechanically.

A further source of uncertainty arises from the manual assignment of mountain and valley folds after laser scoring. Small inconsistencies in sharpness can introduce additional variability in the energy landscape and the critical snapping force.

The measurement instruments also contribute a quantifiable level of error. The Torbal FB5 force sensor has a manufacturer-specified uncertainty of ±0.0002~lbF, and the motorized actuator introduces slight positional hysteresis. Together, these correspond to approximately ±1\% uncertainty in the recorded force values, affecting the absolute magnitudes but not the observed trends.

Additionally, alignment imperfections of the origami specimen within the supporting frame or friction at contact points can cause asymmetric loading, which in turn distorts the measured force–displacement curve near the snap transition.

Overall, these uncertainties do not alter the principal findings of this study---the existence of a tunable, geometry-dependent snapping transition in the Mars pattern---but they define the realistic precision limits of the measurements reported here.

\section{Conclusions}

\begin{enumerate}

    \item The Mars tessellation (with the studied MV assignment) is not rigidly foldable. Folding-speed compatibility for degree-4 vertices shows that the required major/minor speed ratios cannot be propagated consistently across neighboring units. This global kinematic incompatibility precludes any continuous rigid-folding motion with flat facets, even if diagonal creases are added to the square facets.

    \item In a thin elastic realization, this incompatibility is resolved through facet bending, which stores elastic energy and produces a snap-through discontinuity in the force--displacement response. For the baseline \(3\times3\) pattern (\(N=31\)), the dominant snap is highly reproducible, with a mean force drop of \(92.6 \pm 5.5\%\) at \(\bar{x}_{\mathrm{snap}} = -0.039 \pm 0.002~\mathrm{m}\).

    \item Increasing system size changes the character of the instability. 
    The \(3\times3\) tessellation exhibits a single global snap event, whereas the \(7\times7\) pattern displays multiple discrete force drops, indicating that incompatibility is relieved through spatially localized snap-through transitions in extended sheets.

    \item The snap magnitude is tunable. 
    Adding diagonal creases and increasing their scoring depth via laser power systematically reduces the force drop at the dominant snap, demonstrating that controlled crease compliance lowers the effective energy barrier and suppresses the instability.
    
\end{enumerate}

Greater testing samples are required for more robust conclusions, particularly with respect to the control of snap-through behavior through the introduction of additional creases with modified scoring depths. Establishing such robustness would enable systematic and reliable tuning of the magnitude of snap-through instabilities, which is essential for translating geometrically frustrated origami patterns into practical mechanical metamaterials and could be exploited in applications such as impact- or energy-absorbing components.

For testing the snap-through behavior in larger patterns (e.g., the \(7\times7\) tessellation examined in Section~\ref{exp1_results}), high-speed video or marker tracking could be used in future studies to determine whether the final snap involves a larger spatial fraction of the tessellation than earlier snap events. Such measurements would help substantiate the observation that the final snap exhibits the largest force drop and would further support the physical interpretation proposed in this work.

In addition, future work could employ origami simulators similar to the one used in this study to identify optimal locations for additional diagonal creases, allowing for a direct comparison between simulated strain distributions and experimental measurements. Further improvements include testing different sheet materials, such as papers with varying thickness or stiffness, to assess the robustness of the observed behavior across material properties.

\section*{Acknowledgments}

\noindent\textbf{Funding} The authors gratefully acknowledge the support of the U.S. National Science Foundation under grant DMS-2347000, ``RUI: Configuration Spaces of Flexible Polyhedral Surfaces.''

\vspace{0.5em}

\noindent\textbf{Institutional Support} The authors thank the Department of Physics and Astronomy at Franklin \& Marshall College for providing laboratory space and experimental equipment used in this study.

\vspace{0.5em}

\noindent\textbf{Data Availability} The data and code supporting the findings of this study are publicly available on GitHub at this repository: \href{https://github.com/menelrap/Tunable-Snapping-and-Rigid-Foldability-in-the-Mars-Origami-Pattern}{https://github.com/menelrap/Tunable-Snapping-and-Rigid-Foldability-in-the-Mars-Origami-Pattern}.

\vspace{0.5em}

\noindent\textbf{Competing Interests} The authors declare no competing financial or non-financial interests.

\onecolumn

\appendix
\section{Infinitesimal Rigid-Foldability Analysis of the Modified Mars Tessellation}
\label{app:rigid_analysis}

In this Appendix we provide some details on the rigid foldability of the Mars tessellation MV assignment studied in the present work when additional diagonal creases are added across the square faces of the crease pattern. 

As shown in Section~\ref{sec:rigid_foldability}, our Mars MV assignment is not rigidly foldable on its own. However, adding diagonal creases to the square faces will change the vertices from degree-4 to degree-5 or -6, thus increasing the degrees of freedom of the vertices. This alters the origami kinematics greatly and could allow it to be rigidly foldable. However, this is not the case; the Mars pattern with the MV assignment considered in this paper remains non-rigidly foldable even if we add diagonals without a specified MV assignment to the square faces.  

\begin{figure}[h!]
    \centering
    \includegraphics[width=.5\textwidth]{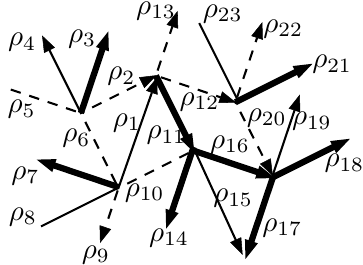}
    \caption{A section of the Mars pattern with our MV assignment and diagonals added to the square faces.  Dashed lines are valleys, bold lines are mountains, and non-dashed, non-bold lines can be mountain or valley. Arrows indicate which vertices determine folding angles and propagate them along the crease to a neighboring vertex.}

    \label{fig:adding_more_creases2}
\end{figure}

To see this, consider the patch of the Mars tessellation with square diagonals added as shown in Figure~\ref{fig:adding_more_creases2}. In general, an origami crease pattern on a simply connected piece of paper with $E$ crease lines and $V$ internal vertices will have $E-3V$ degrees of freedom as a rigid folding mechanism \citep{Tachi2}. The Mars patch in Figure~\ref{fig:adding_more_creases2} has $E-3V=23-18=5$, which means we should be able to have five of the creases have independent folding angles (aside, perhaps, from their MV assignment) and all the remaining folding angles be determined. In Figure~\ref{fig:adding_more_creases2} we let $\rho_5, \rho_6$ be independent at the left-most vertex, which then determines $\rho_2, \rho_3, \rho_4$ and we drew arrows on their respective creases to indicate how they would serve as inputs to other vertices. We then proceeded left-to-right, so that in the next vertex, which is degree-6, $\rho_6$ is already independent, and we chose $\rho_8$ and $\rho_{10}$ to also be independent. This determines the folding angles for $\rho_7, \rho_9$, and $\rho_1$, and so on.

To calculate the dependence of the determined folding angles from the independent ones, we calculated their infinitesimal folding angle relationships as described in \cite{WatanabeKawaguchi2009}, where matrix equations are given that follow from the basic matrix model of rigid origami \cite{BelcastroHull2002}. Specifically, we describe a vertex with $n$ creases in an origami crease pattern by the angles $\theta_i =$ the angle of the crease measured from the positive $x$-axis. If we let $\rho_i$ denote a small change in the folding angle of the $i$th crease from the unfolded state, then the first-order constraints of an infinitesimal motion from the unfolded state can be captured by
\begin{equation}\label{eq:matrix1}
    A\vec{\rho} = \begin{pmatrix}
        \cos\theta_1 & \cos\theta_2 & \cdots & \cos\theta_n\\
        \sin\theta_1 & \sin\theta_2 & \cdots & \sin\theta_n
    \end{pmatrix}
    \begin{pmatrix}
        \rho_1\\
        \rho_2\\
        \vdots\\
        \rho_n
    \end{pmatrix}=\vec{0},
\end{equation}
and the second-order constraints can be described by 
\begin{equation}\label{eq:matrix2}
{\vec{\rho}}^T C \vec{\rho}=0\mbox{ with }C=
\begin{pmatrix}
    0 & c_{1,2} & c_{1,3} & \cdots & c_{1,n}\\
    c_{1,2} & 0 & c_{2,3} & \cdots & c_{2,n}\\
    c_{1,3} & c_{2,3} & 0 & \cdots  & c_{3,n} \\
    &  \vdots & & \ddots & \vdots \\
    c_{1,n} & c_{2,n} & c_{3,n} & \cdots & 0
\end{pmatrix}
\end{equation}
where $c_{i,j}=\cos\theta_i\sin\theta_j-\sin\theta_i\cos\theta_j$.

For example, at the first, left-most vertex in Figure~\ref{fig:adding_more_creases2}, we may let the $\rho_2$ crease be the $x$-axis, and so our angles $\theta_1,\ldots, \theta_5$ are $0^\circ, 45^\circ, 90^\circ, 135^\circ$, and $270^\circ$. Then solving equations \eqref{eq:matrix1} and \eqref{eq:matrix2} for $\rho_2$, $\rho_3$, and $\rho_4$ in terms of the independent variables $\rho_5$ and $\rho_6$ gives two sets of solutions, one of which is not compatible with our chosen MV assignment. This gives
\begin{eqnarray*}
   \rho_2 & = & \frac{1}{2}\left((-1+\sqrt{2})\rho_5+\frac{\sqrt{-2\rho_5 K}}{\sqrt{2}}\right)\\
   \rho_3 & = & \frac{1}{2}\left(\sqrt{2}\rho_5-\sqrt{-2\rho_5 K}\right)\\
   \rho_4 & = & \frac{1}{4}\left((-4-\sqrt{2})\rho_5+4\rho_6+2\sqrt{-\rho_5K}-\sqrt{-2\rho_5K}\right)
\end{eqnarray*}
where $K=(1+2\sqrt{2})\rho_5-4\sqrt{2}\rho_6$. Note that the way to interpret these equations is that they represent the folding angle dependencies when folding the creases a small amount away from the unfolded (folding angles equal zero) state. However, these equations need to give real values, and the MV assignment must be satisfied (meaning that $\rho_2, \rho_4,\rho_5, \rho_6 >0$ and $\rho_3<0$). Simplifying the equations with these constraints we may obtain the condition
$$\rho_6>\left(\frac{3\sqrt{2}}{2}-1\right)\rho_5>0.$$
Continuing with the other vertices we may obtain the following equations and inequalities:
\begin{eqnarray*}
    \rho_{10} & > & \frac{1}{2}((4-\sqrt{2})\rho_6+4\rho_8)\\
    \rho_{11} & < & -\frac{2+\sqrt{2}}{2}\rho_{10}\\
    \rho_1 & = & \frac{1}{4}(2\rho_{10}-\sqrt{2}\rho_6+\sqrt{L})\\
    \rho_{11} & = & -\rho_1+\rho_2-\sqrt{(1-2\sqrt{2})\rho_1^2+(2\sqrt{2}-2)\rho_1\rho_2+2\rho_2^2}
\end{eqnarray*}
where $L=4(1+\sqrt{2})\rho_{10}^2+4\sqrt{2}\rho_{10}\rho_6+(2-4\sqrt{2})\rho_6^2-4\sqrt{2}\rho_8^2$. The square diagonal crease with folding angle $\rho_1$ also has restrictions, but they depend on whether the crease is a mountain or a valley. Specifically, we obtain:
$$\mbox{if }\rho_1>0\mbox{ then }\rho_2>\frac{1}{2}\left(1-\sqrt{2}+\sqrt{1+2\sqrt{2}}\right)\rho_1.$$
Experimenting with this set of inequalities numerically reveals that there is no simultaneous solution satisfying all of these inequalities.

The fact that the crease pattern in Figure~\ref{fig:adding_more_creases2} is not rigidly foldable, and thus neither is our special MV-assigned Mars pattern with square diagonal creases, may seem surprising. After all, Figure~\ref{fig:adding_more_creases2}'s crease pattern has five degrees of freedom, which should be enough to allow it to rigidly fold. To this we make two points by way of explanation. 

First, a normal degree-of-freedom calculation does not take specific MV assignments into consideration, and it is difficult to see how the imposition of  mountain and valley directions on most of the creases in Figure~\ref{fig:adding_more_creases2} will impact the degrees of freedom.

Second, previous studies have shown that when trying to rigidly fold square twist crease patterns by adding a diagonal crease to the square, the result is often only rigidly foldable by taking a convoluted path in the pattern's configuration space \citep{Hull8,Feng2020}. For example, the crease pattern might need to fold nearly in half along the square diagonal, leaving many creases unfolded for a time, before folding all the creases into their proper configuration. Of course, in the Mars pattern this is not possible, since none of the creases traverse the whole paper. But this prior work on square twists with diagonal creases indicates that folding a square twist tessellation, like Mars, rigidly from either the unfolded state or the fully flat-folded state, could be impossible to do if it means some creases have to fold much faster than others, leading to snap-through instabilities even in the presence of more degrees of freedom.

\twocolumn

\bibliography{references}
\end{document}